\documentclass[aps,prl,showpacs,floatfix,twocolumn,letterpaper,superscriptaddress]{revtex4}
\usepackage[pdftex]{graphicx}
\usepackage{amsmath}
\usepackage{hyperref}
\usepackage{latexsym}
\usepackage{color}
\usepackage{amssymb}
\usepackage{amsmath}

\begin{document}

\title{Anisotropic Polarizability of Ultracold Polar $^{40}$K$^{87}$Rb Molecules}

\author{B. Neyenhuis}
\affiliation{JILA, National Institute of Standards and Technology and University of Colorado, and Department of Physics, University of Colorado, Boulder, Colorado 80309-0440, USA}
\author{B. Yan}
\affiliation{JILA, National Institute of Standards and Technology and University of Colorado, and Department of Physics, University of Colorado, Boulder, Colorado 80309-0440, USA}
\author{S. A. Moses}
\affiliation{JILA, National Institute of Standards and Technology and University of Colorado, and Department of Physics, University of Colorado, Boulder, Colorado 80309-0440, USA}
\author{J. P. Covey}
\affiliation{JILA, National Institute of Standards and Technology and University of Colorado, and Department of Physics, University of Colorado, Boulder, Colorado 80309-0440, USA}
\author{A. Chotia}
\affiliation{JILA, National Institute of Standards and Technology and University of Colorado, and Department of Physics, University of Colorado, Boulder, Colorado 80309-0440, USA}
\author{A. Petrov}
\altaffiliation{Alternative address: St. Petersburg Nuclear Physics
Institute, Gatchina, 188300; Division of
Quantum Mechanics, St.Petersburg State University, 198904, Russia}
\affiliation{Department of Physics, Temple University, Philadelphia, Pennsylvania 19122-6082, USA}
\author{S. Kotochigova}
\affiliation{Department of Physics, Temple University, Philadelphia, Pennsylvania 19122-6082, USA}
\author{J. Ye}
\affiliation{JILA, National Institute of Standards and Technology and University of Colorado, and Department of Physics, University of Colorado, Boulder, Colorado 80309-0440, USA}
\author{D. S. Jin}
\affiliation{JILA, National Institute of Standards and Technology and University of Colorado, and Department of Physics, University of Colorado, Boulder, Colorado 80309-0440, USA}

\begin{abstract}
We report the measurement of the anisotropic AC polarizability of ultracold polar $^{40}$K$^{87}$Rb molecules in the ground and first rotationally excited states.  Theoretical analysis of the polarizability agrees well with experimental findings. Although the polarizability can vary by more than $30\%$, a ``magic'' angle between the laser polarization and the quantization axis is found where the polarizability of the $|N=0,m_N=0\rangle$ and the $|N=1,m_N=0\rangle$ states match.  At this angle, rotational decoherence due to the mismatch in trapping potentials is eliminated, and we observe a sharp increase in the coherence time. This paves the way for precise spectroscopic measurements and coherent manipulations of rotational states as a tool in the creation and probing of novel quantum many-body states of polar molecules.
\end{abstract}


\maketitle

The creation of a gas of ultracold polar molecules with a high phase space density \cite{Ni.Science.2008} brings new possibilities beyond experiments with ultracold atomic gases.  In particular, long-range, anisotropic, and tunable dipole-dipole interactions open the way for novel quantum gases, with applications including strongly correlated many-body systems \cite{Baranov200871,Pupillo,Baranov.PhysRevA.83.043602,goral.PhysRevLett.88.170406}, precision measurement \cite{Hudson.PhysRevLett.89.023003}, and ultracold chemistry \cite{Krems.PCCP.2008,Ospelkaus.Science.2010}. Molecules also have complex internal structure with many more internal degrees of freedom than atoms.  In particular, the rotational degree of freedom provides a set of long-lived excited states that are easily coupled to the ground state with microwaves.  Because of the accessible frequency in the microwave domain and the narrow intrinsic linewidth, the transition between rotational states could be ideal for use as a spectroscopic probe of the system.  For example, such a narrow transition could be used to measure small energy shifts due to dipolar interactions \cite{Hazzard_PhysRevA.84.033608}. In addition, dipole-dipole interactions can be realized without applying a DC electric field but instead by directly coupling the two lowest rotational states with a microwave field. Within the rotating frame of the microwave transition, there is a strong dipolar interaction, which can be used to model novel quantum many-body Hamiltonians \cite{Goshkov_PhysRevLett.107.115301}.  Additionally, the microwave detuning and power can be varied to modify collision dynamics \cite{Micheli.PhysRevA.76.043604}. This has been proposed as a way to achieve a topological superfluid of paired fermionic polar molecules \cite{Cooper.PhysRevA.84.013603}.

A prerequisite for such experiments is long coherence times for the interaction between a microwave field and the rotational states.  However, for molecules confined in an optical dipole trap, the difference in AC (or dynamic) polarizability between different rotational states must be considered \cite{Ospelkaus.FaradayDiscussion}.  A difference in polarizability leads to different trap frequencies and spatially dependent variations in the rotational transition frequency, which can lead to dephasing and decoherence \cite{Ye.Science.2008.magicwavelength}. For atoms, if the trapping light is far detuned compared to the energy splitting between two states, their AC polarizabilities will be nearly equal. In contrast, for molecules, the ground and the first rotationally excited state with the same angular momentum projection onto the quantization axis have different parity and will therefore couple to different electronic excited states. This can result in AC polarizabilities that differ by more than 30\%, even when the light is far detuned \cite{Kotochigova.PhysRevA.82.063421}.

In atomic systems, it is possible to adjust the wavelength of the trapping light such that the polarizabilities of two states of interest (often clock states in alkaline earth atoms) are the same \cite{Ye.Science.2008.magicwavelength}.  Although, in principle, one could find such a ``magic'' wavelength trap for molecules \cite{Zelevinsky.PhysRevLett.100.043201}, the large number of additional states from rotation and vibration makes it difficult to find a suitable wavelength that is sufficiently detuned such that off-resonant light scattering is negligible.  However, molecules provide a different way to adjust the polarizability; the AC polarizability of a molecule depends on the relative orientation of the molecule and the polarization of the trapping light \cite{Kotochigova.PhysRevA.82.063421}.

In this Letter, we explore the interaction between the trapping light and the molecules by examining the real part of the AC polarizability. (The imaginary part of the polarizability was reported in a previous study of the lifetime of molecules trapped in a three-dimensional lattice~\cite{Chotia.PhysRevLett.108.080405}.) In particular, we determine how the polarizability of the rotationally excited states depends on the relative orientation of the quantization axis, $\hat{z}$ (which in these experiments is given by a magnetic field of 545.9 G), and the polarization of the trapping light.  A ``magic'' angle exists where the polarizability of the $|N=0,m_N=0\rangle$ (where $N$ is the rotation quantum number, and $m_N$ is its projection onto the quantization axis) and $|N=1,m_N=0\rangle$ states match, making the AC Stark shift the same for these two internal states of the molecule.  This ``magic'' angle is expected at $\cos^2\theta=1/3$, or $\theta \approx54$ degrees \cite{demille_budker}. We observe this magic angle both through direct measurements of the AC polarizability and through measurements of the coherence time for driving the $|0,0\rangle$ to $|1,0\rangle$ transition.  In the theory component of this Letter, we extend the ideas of mixing of rotational levels due to static electric fields of Ref.~\cite{Kotochigova.PhysRevA.82.063421} to include mixing due to the intrinsic nuclear electric-quadrupole and Zeeman interactions discussed in Ref.~\cite{Ospelkaus.PhysRevLett.104.030402}. We compare our experimental results to the analytic results of an approximate Hamiltonian, which mixes the three projections of the first rotationally excited state.

\begin{figure}
    \includegraphics[height=5cm]{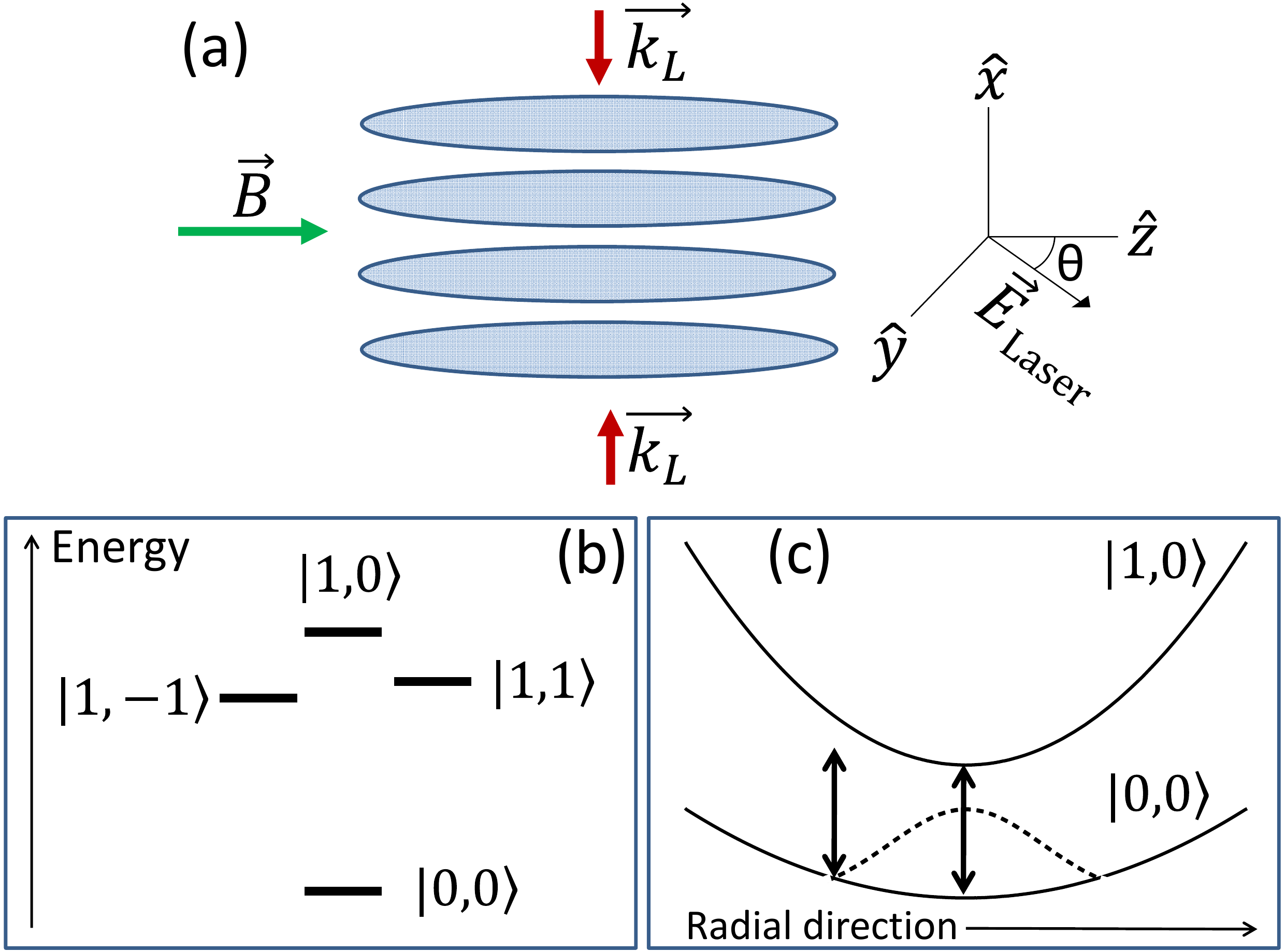}
    \caption{ (a) Experimental schematic. The lattice beam propagates along $\hat{x}$, the magnetic field points in the $\hat{z}$ direction, and the polarization of the lattice light makes an angle $\theta$ with the magnetic field in the y-z plane.  (b) Schematic of rotational energy states.  The degeneracy of the $N=1$ level is split in a magnetic field.  (c) A sketch of the optical dipole potentials for the $|0,0\rangle$ and $|1,0\rangle$ states.  A Gaussian is overlaid to show the density distribution of the molecular cloud in the trap.  When the two states are connected by a 2.22 GHz microwave drive, there is effectively a spatially varying detuning across the cloud due to the difference in the trap potentials.}
    \label{fig:traps}
\end{figure}

Molecules have a definite orientation that strongly affects their polarizability.  In the frame of a diatomic molecule such as KRb, the DC (or far-off-resonance AC) polarizability along the internuclear axis is much stronger than the polarizability perpendicular to the internuclear axis.  This has been successfully exploited to align molecules with intense laser pulses in order to study the stereodynamics of chemical reactions \cite{Herschbach.PhysRevLett.74.4623,seideman:7887,sakai.molecularalignment}.  The $N=0$ rotational ground state is spherically symmetric and therefore its AC polarizability has no dependence on the relative orientation of the laser polarization and the quantization axis. However, the rotational wavefunctions of the three projections ($m_N$) of the $N=1$ rotationally excited state have well-defined orientations relative to the quantization axis. For example, the $|1,0\rangle$ state corresponds to a $p_z$ orbital that is aligned along the $\hat{z}$-axis, and therefore the polarizability is the largest when the polarization of the AC field is along the $\hat{z}$-axis.

The degeneracy of the three projections of the $N=1$ state is broken by the hyperfine interaction, specifically, the interaction between the nuclear quadrupole moment and the rotation of the molecule~\cite{Ospelkaus.PhysRevLett.104.030402,Hutson_PhysRevA.80.043410}.  In the rotationally excited states, this coupling gives a specific nuclear spin state up to a $5\%$ admixture of other hyperfine states; however, we will ignore these contributions and will work only with states whose dominant nuclear spin character is $m_I^{\mathrm{K}}=-4,m_I^{\mathrm{Rb}}=1/2$ unless otherwise noted, where $m_I^{\mathrm{K}}$ and $m_I^{\mathrm{Rb}}$ are the nuclear spin projections onto the $\hat{z}$ axis. For more discussion of the nuclear quadrupole couplings, see the online supplementary material. At a field of 545.9 G and with no light, we find that the $|1,1\rangle$ state is 58 kHz above the $|1,-1\rangle$ state and the $|1,0\rangle$ state is 268 kHz above the $|1,1\rangle$ state (Fig \ref{fig:traps}b).  The splitting between the $m_N = \pm$ 1 states is of the same order of magnitude as the AC Stark shifts induced by the trapping lasers. Consequently, we calculate the anglar dependence of the dynamic polarizability of the three rotationally excited states using perturbation theory that includes the light-induced couplings between the bare states (no light).

Using the general formalism of Ref.~\cite{Bonin}, the complex dynamic polarizability for the $|0,0\rangle$ state is given by
\begin{equation}
   \alpha_{|0,0\rangle} = \frac{1}{3}(\alpha_{\parallel} + 2\alpha_{\perp})
   \label{m0}
\end{equation}
where the ``reduced'' polarizabilities $\alpha_{\parallel}$ and $\alpha_{\perp}$ are the parallel and perpendicular (with respect to the intermolecular axis) polarizabilities that describe the averaged contributions from ro-vibrational states of all electronically excited $^1\Sigma^+$ and $^1\Pi$ potentials. As mentioned above, the polarizability of the $|0,0\rangle$ state is independent of the angle $\theta$ between $\hat{z}$ and the polarization of the 1D optical lattice used to trap the molecules (see Fig.~\ref{fig:traps}a).

For the $|1,0\rangle$ and $|1,\pm1\rangle$ states, the dressed or mixed polarizabilities at laser intensity $I$ are given by the total Stark shift divided by the total intensity, $\alpha_{j}=-\frac{E_{j}(I)-E_{j}(0)}{I}$, where $E_{j}$, with $j=1,2,3$, are the  eigenvalues of the $3\times3$ Hamiltonian
\begin{eqnarray*}
&&    \begin{array}{ccc}
     \quad \quad   | 1,0\rangle &  \quad\quad|1,-1\rangle & \quad\quad |1,1\rangle
     \end{array}\\
H &=&
      \left(
    \begin{array}{ccc}
-\alpha_{11}I + \epsilon_1 & -\alpha_{12}I &  -\alpha_{13}I \\
-\alpha_{12}I & -\alpha_{22}I + \epsilon_2 &  -\alpha_{23}I \\
-\alpha_{13}I & -\alpha_{23}I &  -\alpha_{33}I + \epsilon_3
    \end{array}
    \right).
\end{eqnarray*}
Here,
\begin{eqnarray*}
 \alpha_{11}& = & \frac{\alpha_{\parallel}+4\alpha_{\perp}}{5} \sin^2\theta +
\frac{3\alpha_{\parallel}+2\alpha_{\perp}}{5} \cos^2\theta\\
\alpha_{22} &= & \alpha_{33} 
  =  \frac{2\alpha_{\parallel}+3\alpha_{\perp}}{5} \sin^2\theta +
\frac{\alpha_{\parallel}+4\alpha_{\perp}}{5}\cos^2\theta \nonumber \\
\alpha_{12}& =&-\alpha_{13}  = \sqrt{2}\,\frac{ \alpha_{\parallel} - \alpha_{\perp} }{5}\sin\theta \cos\theta\\
\alpha_{23} &=& \frac{1}{5}(\alpha_{\perp} - \alpha_{\parallel})\sin^2\theta \,,
\end{eqnarray*}
and $\epsilon_1$, $\epsilon_2$, and $\epsilon_3$ are the energies for states $|1,0\rangle$, $|1,1\rangle$ and $|1,-1\rangle$, respectively, at $I=0$.

\begin{figure}
    \includegraphics[height=6cm]{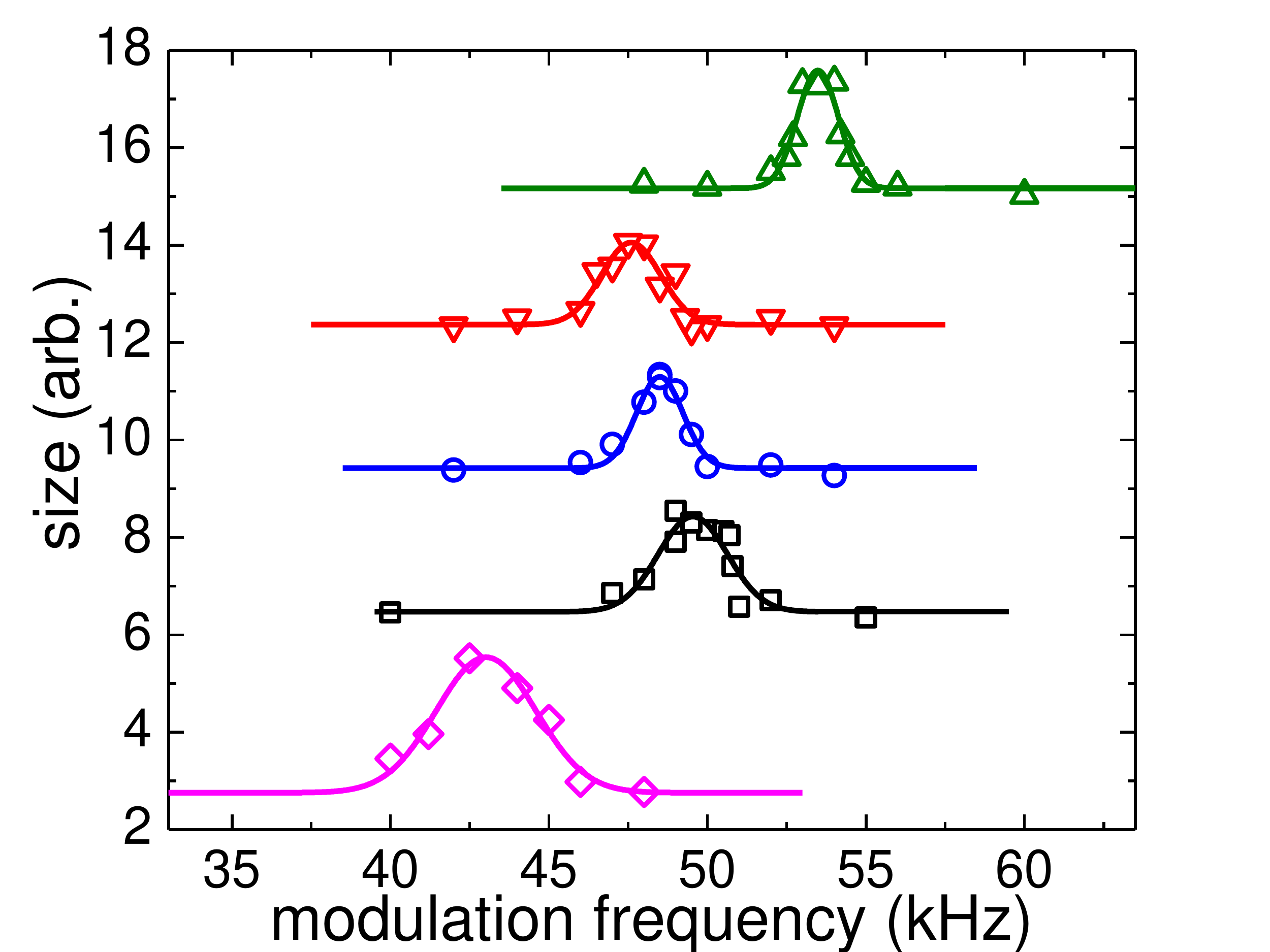}
    \caption{Parametric heating resonances in the far-off-resonance optical dipole trap for $\theta=57$ degrees. The y-axis shows the rms size in $\hat{x}$ of an expanded gas of KRb (Rb) after 5 ms (21 ms) of time of flight. The curves have been offset vertically for clarity. Using Gaussian fits (lines), we determine the center of the parametric heating resonances for (from bottom to top) Rb, and KRb in the $|0,0\rangle$, $|1,0\rangle$, $|1,1\rangle$, and $|1,-1\rangle$ states. The resonant frequency allows us to extract the trap depth for each state.}
    \label{fig:parametricheating}
\end{figure}

We measure the AC polarizability of the molecules in a one-dimensional optical lattice with a peak intensity, $I_0$, of 2.3 kW/cm$^2$ and a wavelength of $\lambda =$ 1064 nm \cite{Miranda.NaturePhys.2011,Chotia.PhysRevLett.108.080405}. Using a single microwave pulse, we can selectively transfer population from the rotational ground state, $|0,0\rangle$, to any of the projections of the $N=1$ state with near $100\%$ efficiency \cite{Ospelkaus.PhysRevLett.104.030402}.  We measure the trap depth, $U_{\mathrm{KRb}}$, from which we can extract the AC polarizability, by measuring the parametric heating resonance (see Fig.~\ref{fig:parametricheating}). We modulate the intensity of the optical lattice for 4 ms with an amplitude of approximately 10\% of the total depth. The modulation frequency is varied to find the resonant frequency where molecules are excited from the lowest band of the lattice to the second excited band.  In the deep lattice limit, this resonant frequency is twice the trap frequency. However, we operate our lattice in an intermediate intensity regime where the relationship between the trap depth and the resonant frequency must be extracted from a numerical solution to the lattice potential.  We determine the polarizability, without having to characterize the optical beam parameters such as power or beam waist, by comparing the molecular results to a similar measurement of the trap depth for Rb.  The polarizability of KRb is then given by $\alpha_{\textrm{KRb}} = \alpha_{\textrm{Rb}} U_{\textrm{KRb}}/U_{\textrm{Rb}}$, where $\alpha_{\textrm{Rb}}/h=3.242\times10^{-5}$ MHz/(W/cm$^2$) at 1064 nm \cite{Safronova_PhysRevA.69.022509}, and $h$ is Planck's constant.

\begin{figure}
    \includegraphics[height=6cm]{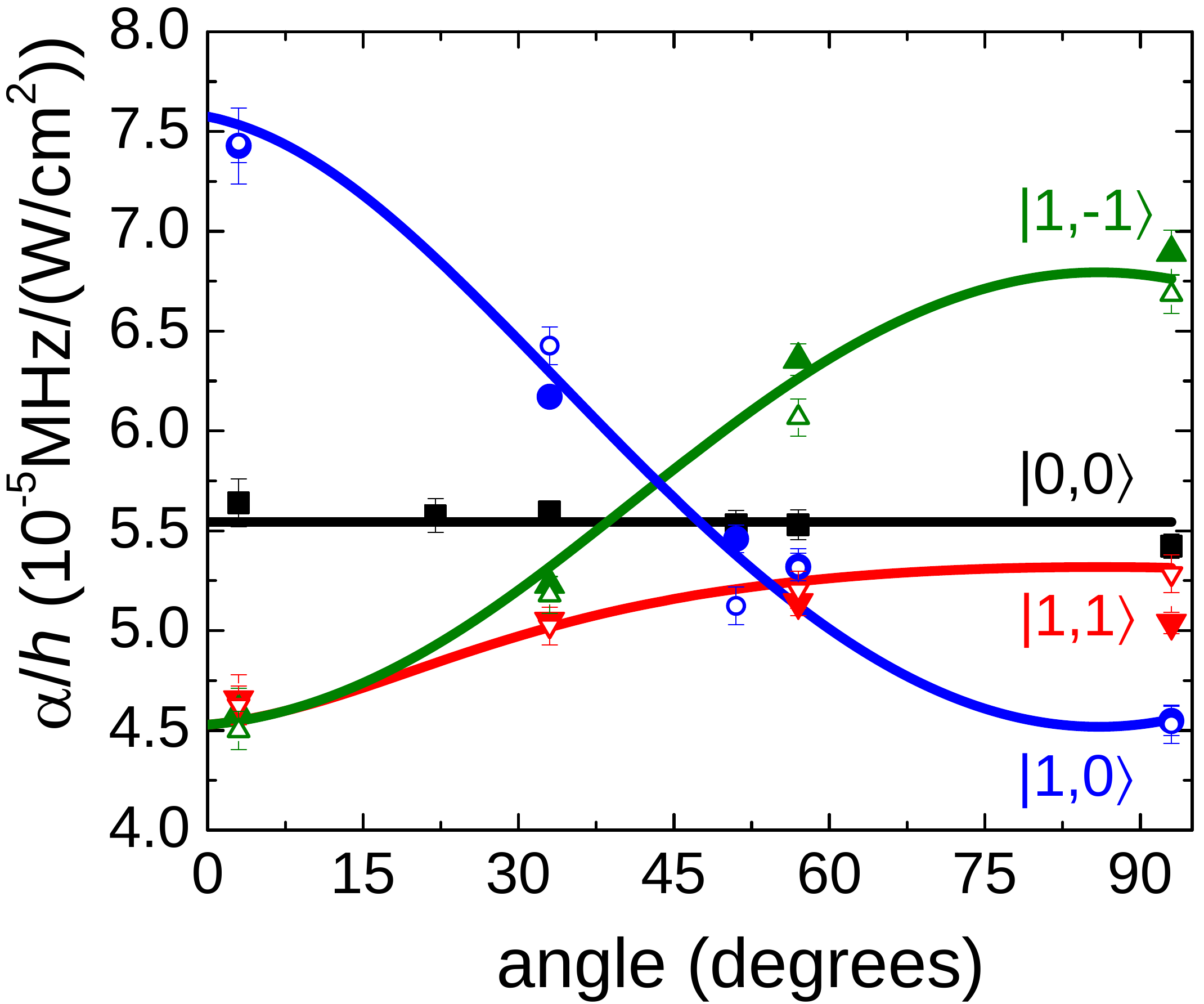}
    \caption{The AC polarizability of KRb at 1064 nm for the $|0,0\rangle$ (black squares), $|1,0\rangle$ (blue circles), $|1,1\rangle$ (red inverted triangles), and $|1,-1\rangle$ (green triangles) states. Error bars are from the fit uncertainty in the center of the parametric heating resonances and correspond to $\pm1$ standard deviation.  Theory lines are a simultaneous fit to Eqn.~\ref{m0} and the polarizabilities from the solution of $H$ with three free parameters $\theta_m$, $\alpha_\perp$, and $\alpha_\parallel$. Open circles represent a separate measurement where the polarizability is extracted from the shift in the microwave transition frequency.}
    \label{fig:angle}
\end{figure}

In Fig.~\ref{fig:angle}, we show the polarizabilities of the $N=0$ ground state as well as the three projections of the $N=1$ rotationally excited state as a function of $\theta$, which is varied by adjusting a half-waveplate in the lattice beam.  Although the rotation of the half-waveplate allows us to choose $\theta$ with precision better than one degree, the absolute alignment relative to the magnetic field has an estimated systematic uncertainty of $\pm3$ degrees.  We fit Eqn.~\ref{m0} and the polarizabilities from the eigenenergies of $H$ to the experimental data with three free parameters, $\theta_m$, $\alpha_\parallel$ and $\alpha_\perp$.  From the best fit, we determine that the ``magic'' angle $\theta_m=48(4)$ degrees, $\alpha_\parallel/h =10.0(3) \times 10^{-5}$ MHz/(W/cm$^2$), and $\alpha_\perp/h=3.3(1) \times 10^{-5}$ MHz/(W/cm$^2$).

Hyperfine couplings between the $|1,0\rangle$ and $|1,\pm1\rangle$ states result in a small change of the predicted magic angle from 54 degrees.  For the polarizabilities and intensity given above, we expect $\theta_m = 52$ degrees, which agrees with our measurement to within the error. For comparison to the measurement, theoretical values of the polarizabilities  $\alpha_{\parallel}$ and $\alpha_{\perp}$ are obtained using non-relativistic potentials and dipole moments described in Ref.~\cite{Kotochigova.PhysRevA.82.063421}, and are $h \times 12.2 \times 10^{-5}$ MHz/(W/cm$^2$) and $h \times 2.01 \times 10^{-5}$ MHz/(W/cm$^2$), respectively.  In principle, $\alpha_{\parallel}$ and $\alpha_{\perp}$ depend on the ro-vibrational state of the molecule, however, we find in the calculation that for small $N$, the polarizabilities $\alpha_{\parallel}$ and $\alpha_{\perp}$ are independent of $N$ to better than 0.01\%.

Because the bare states are mixed by the lattice light, the dresssed state polarizabilities depend on the intensity of the light.  This is effect is strongest for the $|1,\pm1\rangle$ states, where the energy splitting of the bare states is relatively small. However, we experimentally verify that at the intensity used here this effect is small.  For example, for a $50\%$ increase in intensity with $\theta=93$ degrees, we see only a $3.9(8)\%$ decrease in the polarizability of the $|1,1\rangle$ state.

We can also use an independent measurement of the shift in the microwave transition frequency to determine the polarizability as a function of $\theta$.
The difference in polarizability between the $|0,0\rangle$ and $|1,0\rangle$ states results in a shift in the microwave transition frequency:
\begin{equation}
f = f_0 + (\alpha_{|0,0\rangle}-\alpha_{|1,0\rangle}) I_0/h + \frac{\omega_{|1,0\rangle}}{4 \pi}-\frac{\omega_{|0,0\rangle}}{4 \pi} +  \Delta f,
\label{eq:shift}
\end{equation}
where $f$ is the measured transition frequency, $f_0$ is the transition frequency from $|0,0\rangle$ to $|1,0\rangle$ with $I=0$ (measured after release from the lattice), $\omega/(4 \pi)=\sqrt{\frac{\alpha I_0}{2 \lambda^2 m}}$ is the trap zero point energy in the lattice divided by $h$, $m$ is the mass, $\Delta f= \frac{\alpha_{|1,0\rangle} - \alpha_{|0,0\rangle}}{\alpha_{|0,0\rangle}} k_B T/h $ is the shift in the center of the transition frequency caused by the spatially dependent detuning, $k_B$ is Boltzmann's constant, and $T=400$ nK is the temperature. The transition frequency is only sensitive to the polarizability difference between the $|0,0\rangle$ and $|1,0\rangle$ states, but can be compared with the polarizability from the direct measurement of the trap depth by fixing $\alpha_{|0,0\rangle}$ and then solving Eqn.~\ref{eq:shift} for $\alpha_{|1,0\rangle}$. The resultant polarizabilities are shown in open symbols in Fig.~\ref{fig:angle}.  We see a good agreement between the two methods.

\begin{figure}
    \includegraphics[height=6cm]{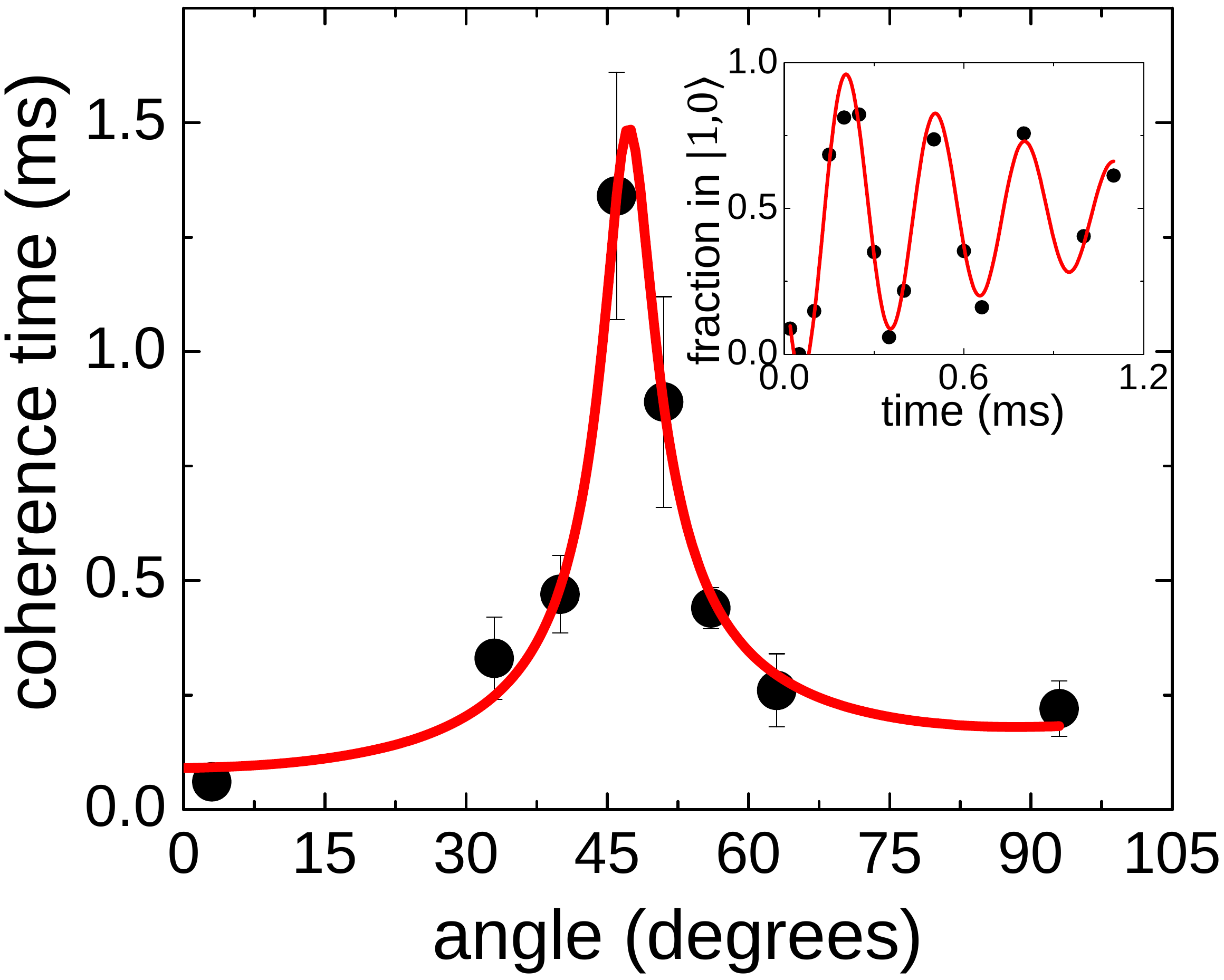}
    \caption{The Ramsey coherence time measured in the one-dimensional optical lattice as a function of angle.  A sharp increase in coherence time is observed at the ``magic'' angle where the polarizabilities of the $|0,0\rangle$ and $|1,0\rangle$ states are matched. Inset: A Ramsey oscillation fit to a damped sine wave to extract the coherence time for $\theta=51$ degrees.}
    \label{fig:decoherence}
\end{figure}

To study the effect of the magic angle, we measure the rotational excitation coherence time as a function of angle, and the results are shown in Fig.~\ref{fig:decoherence}.  We measured the coherence time between the $|N=0,m_N=0,m_I^{\mathrm{K}}=-4,m_I^{\mathrm{Rb}}=1/2\rangle$ and $|N=1,m_N=0,m_I^{\mathrm{K}}=-3,m_I^{\mathrm{Rb}}=1/2\rangle$ states with Ramsey spectroscopy. The probe pulse is 40 $\mu$s long, and the probe frequency is detuned from the resonance by 3 to 12 kHz. We fit the Ramsey oscillation as a function of time to a damped sine wave to extract the coherence time (see inset of Fig.~\ref{fig:decoherence}).

The coherence time due to the mismatch in polarizability should scale as one over the difference in $\partial E / \partial I$, which, because of the intensity dependence of the polarizability, is not the same as the difference in polarizability.  Technically, the $\Delta f$ term in Eqn.~\ref{eq:shift} should also use the ``local polarizability'', $\partial E / \partial I$, instead of $\alpha$ because the trapped gas experiences only a small range of intensities. However, in Eqn.~\ref{eq:shift} the difference between $\alpha$ and $\partial E / \partial I$ gives a small correction to a term that accounts for less than $10\%$ of the total frequency shift, and is therefore negligible. On the other hand, because the coherence time depends critically on this polarizability difference, the ``local polarizability'' must be used.

We fit to the data with a simple model that includes the angular dependence:
\begin{equation}
\tau = 1/\sqrt{(1/T_2)^2+\left(\frac{\partial E_{|1,0\rangle} / \partial I - \partial E_{|0,0\rangle} / \partial I}{\partial E_{|0,0\rangle} / \partial I} \Delta E/\hbar \right)^2}
\label{tau},
\end{equation}
where $T_2$ is the coherence time from all sources of decoherence other than the polarizability, $\Delta E$ is the spread of energy across the cloud in the $|0,0\rangle$ state, and $|\partial E_{|0,0\rangle} / \partial I-\partial E_{|1,0\rangle} / \partial I|$ depends upon $\theta$, $\alpha_\parallel$, $\alpha_\perp$, $I$, $\epsilon_1$, $\epsilon_2$, and $\epsilon_3$. Using the measured values of $I$, $\epsilon_1$, $\epsilon_2$, and $\epsilon_3$ and the fitted values of $\alpha_\parallel$ and $\alpha_\perp$ from Fig.~\ref{fig:angle}, we use Eqn.~\ref{tau} and obtain $T_2=1.5(2)$ ms, $\Delta E/\hbar=3.3(4)\times 10^{4}$ s$^{-1}$, and $\theta_m=46.5(5)$ degrees. The expected value of $\theta_m$ for our values of $I$, $\alpha_\perp$, and $\alpha_\parallel$ is 48 degrees, which agrees with the fitted value to within the systematic error. Note that the expected value of $\theta_m$ is different when probing the difference in $\partial E / \partial I$ rather than $\alpha$. Possible sources of $T_2$ range from technical noise to resonant dipole-dipole interactions \cite{Hazzard_PhysRevA.84.033608}, and further study is required to understand this limit to the coherence time.

In conclusion, we have measured the angular dependence of the AC polarizability of ultracold KRb and observed the ``magic'' angle where the polarizability of the $|0,0\rangle$ and $|1,0\rangle$ states match. At this angle, we are able to increase the coherence time between the two rotational states by an order of magnitude. This opens the way for the use of the rotational states for precision spectroscopy as well as the modification of the collision dynamics in optically trapped molecular samples.

\begin{acknowledgments}
We acknowledge financial support from NIST, NSF, DOE, AFOSR, and DARPA. S. A. M. acknowledges funding from the NDSEG. We thank K. R. A. Hazzard, A. M. Rey, J. L. Bohn, and G. Qu\'{e}m\'{e}ner for useful discussions.
\end{acknowledgments}

\vspace{5 cm}
\textbf{Supplementary Information}

As was shown in Ref.~\cite{Ospelkaus.PhysRevLett.sup.104.030402}, the nuclear quadrupole moment couples to the rotation of the molecule and mixes rotationally excited states with different hyperfine character. It is precisely this coupling that breaks the degeneracy of the three projections of the $N=1$ state.  These energy shifts move the magic angle from the zeroth order value of 54 degrees to 52 degrees. However in the main text, other than to include this broken degeneracy, we completely ignored the couplings between different hyperfine states in the perturbation theory calcuations. To justify this assumption we now compare the results from the approximate $3\times3$ Hamiltonian in the main text to an ``exact'' model that includes the nuclear quadrupole couplings.

We extend the ideas of mixing of rotational levels due to a static external electric field of Ref.~\cite{Kotochigova.PhysRevA.sup.82.063421} to include mixing of rotational-hyperfine levels due to the intrinsic nuclear electric-quadrupole interactions introduced in Ref.~\cite{Ospelkaus.PhysRevLett.sup.104.030402}. First, we construct the full Hamiltonian for the rotational-hyperfine levels labeled by $|N,m_N,m_I^{\rm K},m_I^{\rm Rb}\rangle$, where $N$ is the rotation quantum number, $m_N$ is its projection onto the quantization axis, $m_I^{\rm K}$ is the projection of the K nuclear spin onto the magnetic-field axis, and $m_I^{\rm Rb}$ is the projection of the Rb nuclear spin onto the magnetic-field axis. It includes the nuclear Zeeman interaction, $-g_{\rm a} \mu_N \vec{I_{\rm a}}\vec{B}$ for atom $a=$ K or Rb, where $g_{\rm a}$ is nuclear g-factor and $\mu_N$ is the nuclear magneton of atom $a$.  It also includes the nuclear quadrupole interaction, proportional to $(eqQ)_a/(I_a(I_a-1)) \sum_m (-1)^m C_{2m}(\theta,\phi) T_{2,-m}(I_a,I_a)$, with coupling constants $(eqQ)_a$ for each atom $a$ that couples its nuclear spin to rotational states. Here, $C_{lm}(\theta,\phi)$ is a spherical harmonic and $T_{2m}(I_a,I_a)$ is a rank-2 tensor created from the spin $I_a$, $e$ is the proton charge, $q$ is the electric field gradient, and $Q$ the nuclear quadrupole moment.  Finally, we include a polarizability interaction Hamiltonian, $-(\alpha_{\parallel} {\cal O}_{\parallel} + \alpha_{\perp} {\cal O}_{\perp}) I$, with strengths $\alpha_{\parallel}$
and $ \alpha_{\perp}$,  tensor operators ${\cal O}_{\parallel}$ and ${\cal O}_{\perp}$ that depend on light polarization and rotational angular
momentum $\vec N$, and the laser intensity of the trapping light  $I$.

This Hamiltonian couples $(1+3)\times9\times4=144$ channels  $|N,m_N, m_I^{\rm K}, m_I^{\rm Rb}\rangle$ and has four parameters: the quadrupole interaction constants for each of the two atoms and the ``reduced'' polarizabilities $\alpha_{\parallel}$ and $\alpha_{\perp}$, which are  the vibrationally-averaged $v=0$ parallel and perpendicular  polarizabilities that include contributions from all excited $^1\Sigma^+$ and $^1\Pi$ potentials, respectively. We find  transition energies by diagonalizing the Hamiltonian and analyzing its eigenfunctions to connect to the states that have been observed experimentally. Eigenstates are identified by the channel state with the largest contribution. Typically this contribution is more than 90\% and we are justified in labeling eigenstates with only one particular set of nuclear spin projections $|N,m_N, m_I^{\rm K}, m_I^{\rm Rb}\rangle$.

The two quadrupole interaction parameters were estimated in Ref.~\cite{Ospelkaus.PhysRevLett.sup.104.030402} based on measurements of transition frequencies between sub-levels of the $N=0$ and $N=1$ states in the optical dipole trap. However, when these measurements were made, the state-dependent polarizability and the effects of the trapping potential on the transition frequency were not yet understood, and the measurements for the different states were not taken at a constant intensity. Here we improve these constants using new measurements of the transition energies for three hyperfine levels of the $N=1$ rotational state where the effects of the trapping potential have been removed by measuring the transition energy directly after turning off the trapping light. These transition energies are given in Table~\ref{frequency}.

\begin{table}
\caption{The experimentally measured rotational-hyperfine transition frequencies of the lowest vibrational level of the X$^1\Sigma^+$ potential of KRb at zero laser intensity (i.e. without trapping light) and a bias magnetic field with strength $B=545.9$ G.
Transitions start at the $|N=0,m_N=0,m_I^{\rm K}=-4,m_I^{\rm Rb}=1/2\rangle$ state and go to three hyperfine states $|j\rangle$ within the $N=1$ manifold.}
\label{frequency}

\begin{tabular}{c|c}
  $|j\rangle$         &   Frequency (kHz)  \\ \hline
$|1,0,-4,1/2\rangle $   &   2 228 110(1)        \\
$|1,1,-4,1/2\rangle $   &   2 227 842(1)        \\
$|1,-1,-4,1/2\rangle$   &   2 227 784(1)
\end{tabular}
\end{table}

We  optimize the quadrupole interaction constants to fit to the experimental energies of Table~\ref{frequency}. We find that $B_e/h$=1.1139514(5) GHz, $(eqQ)_{\rm K}/h$=0.452(9) MHz, and $(eqQ)_{\rm Rb}/h=-1.308$(9) MHz. Here, $B_e$ is the rotational constant and $h$ is Planck's constant. The first two values are in good agreement with the previous estimate of Ref.~\cite{Ospelkaus.PhysRevLett.sup.104.030402}.  The $(eqQ)_{\rm Rb}$ coupling constant  has changed by $\approx 7\%$.

\begin{figure}
    \includegraphics[height=5cm]{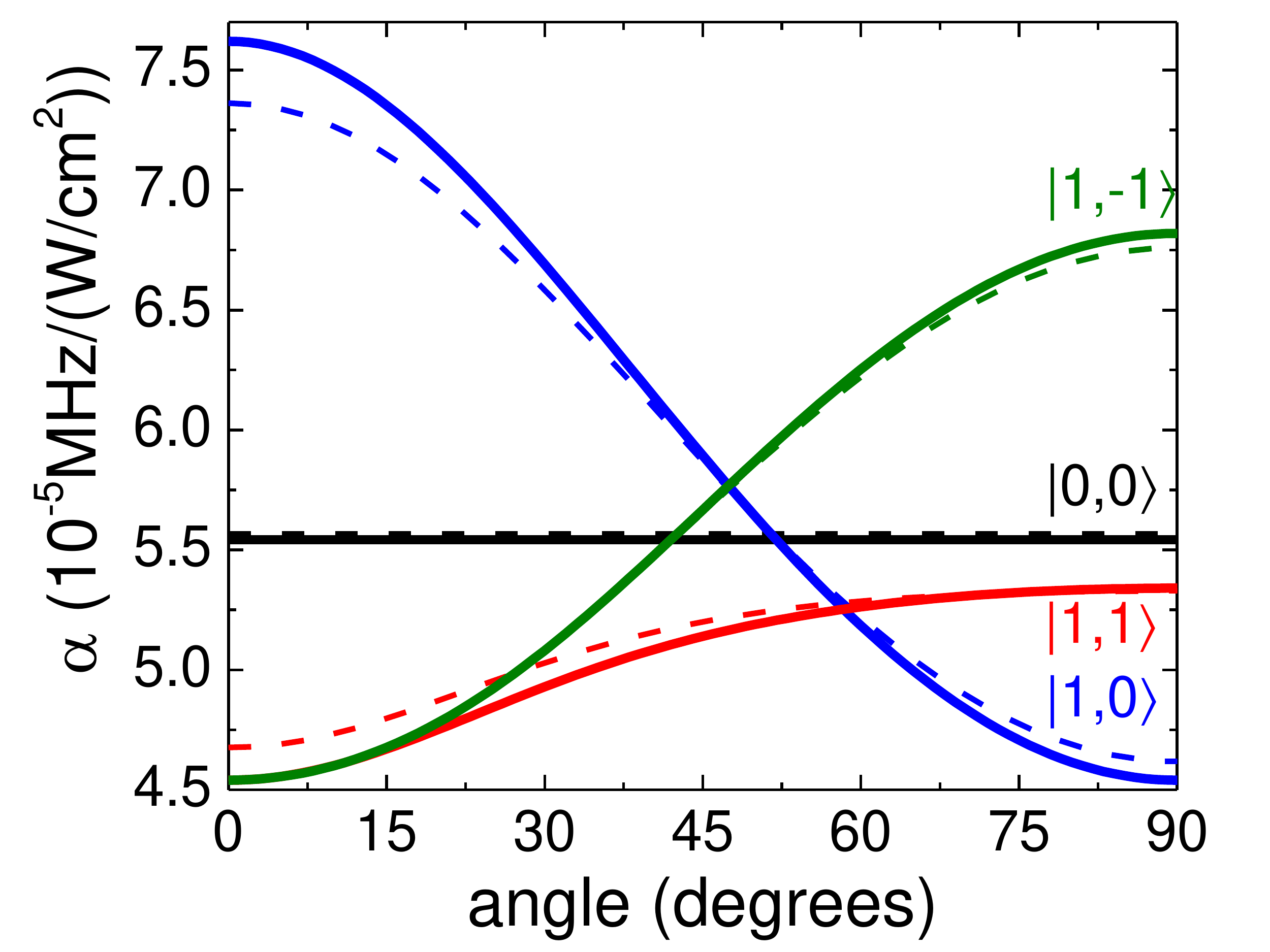}
    \caption{The angle-dependent polarizability for the ``exact'' model (dashed line) and the approximate $3\time3$ Hamiltonian from the main text.}
    \label{fig:exactcomparison}
\end{figure}

The polarizability of eigenstate $j$ with energy $E_j(I)$ is defined as the total Stark shift divided by the total intensity, $\alpha_{j}=-\frac{E_{j}(I)-E_{j}(0)}{I}$.  For comparison to the approximate $3\times3$ Hamiltonian used in the main text, we show the results of the ``exact'' model using the fitted values of $\alpha_\parallel$ and $\alpha_\perp$.  Figure~\ref{fig:exactcomparison} shows the polarizability as a function of $\theta$ based on the ``exact'' Hamiltonian (dashed lines), and the approximate Hamiltonian (solid lines) for the four states studied here and laser intensity $I$=2.35 kW/cm$^2$.  From this comparison, we conclude that the difference between the two models is sufficiently small that the use of the simplified approximate Hamiltonian is justified given the current experimental precision.

\end{document}